\documentclass[namedreferences]{solarphysics}

\usepackage[hyperref,optionalrh]{spr-sola-addons} % For Solar Physics 
\usepackage{graphicx}        % For eps figures, newer & more powerfull
\usepackage{amssymb}        % useful mathematical symbols
\usepackage{color}           % For color text: \color command
\usepackage{breakurl}        % For breaking URLs easily trough lines
\renewcommand{\vec}[1]{{\mathbfit #1}}

\renewcommand{\div}{ \mathrm{div} }
\newcommand{\cur}{ \mathrm{curl} }

\newcommand{\pder}[2]{ \frac{\partial #1}{\partial #2} }
\newcommand{\grad}{ {\bf \nabla } }

\newcommand{\solphys}{{\it Sol. Phys.}}

\chardef\us=`\_

\begin{document}

\begin{article}
\begin{opening}

\title{Thermal Trigger for Solar Flares III: Effect of the
Oblique Layer Fragmentation}

\author[addressref={1},corref,email={leonid.ledentsov@gmail.com}]{\inits{L.S.}\fnm{Leonid}~\lnm{Ledentsov}
\orcid{0000-0002-2701-8871}}
%\author{\inits{}\fnm{}~\lnm{}\orcid{}}
%   NOTE:  Just one corresponding author [corref]
%   \institute{$^{1}$ First affiliation
%                     email: \url{e.mail-a} email: \url{e.mail-b}\\ 
%              $^{2}$ Second affiliation
%                     email: \url{e.mail-c} \\
%             \textit{}
\address[id={1}]{Sternberg Astronomical Institute, 
Moscow State University, 
Moscow 119234, Universitetsky pr., 13, Russia}

\runningauthor{Ledentsov L.S.}
\runningtitle{Thermal Trigger for Solar Flares III}

\begin{abstract}
We consider the oblique fragmentation of the current layer 
as a result of the thermal instability described in 
\citeauthor{2021SoPh..296...74L} (\solphys{} \textbf{296}, 74, 
\citeyear{2021SoPh..296...74L}). 
It is shown that the fragmentation transverse to the current is a natural feature of the model. 
The fragmentation tilt does not exceed a few degrees for realistic preflare parameters of the coronal plasma. 
As a consequence, oblique fragmentation generally does not have a strong impact on the simulation results, 
however, extreme changes can reach an order of magnitude. 
Thus, oblique fragmentation can lead to a decrease in the estimate of the spatial period of the location
of elementary energy release in solar flares to 0.1--1 Mm instead of 1--10 Mm obtained earlier. 
\end{abstract}
\end{opening}
%-------------------------------------------------

\section{Introduction}
     \label{sec1} 

The magnetic reconnection process is a key mechanism 
for changing the topology of the magnetic field \citep{2002A&ARv..10..313P}.
It provides the ability to redistribute independent magnetic fluxes. 
In a highly conductive plasma like the solar atmosphere, 
the reconnection process contains an intermediate stage 
characterized by the formation of a current layer between the interacting magnetic fluxes \citep{1971JETP...33..933S}. 
The current layer, shielding the approaching flows, does not allow them to reconnect. 
In the vicinity of the current layer, the free energy of the magnetic field is accumulated, 
which can be converted into the kinetic energy of plasma particles and electromagnetic radiation
as a result of the destruction of the current layer \citep{2011LRSP....8....6S}. 
We call such an electromagnetic explosion in the solar atmosphere a solar flare \citep{2017LRSP...14....2B}.

It is assumed that the process of destruction of the current layer 
can be associated with various plasma instabilities. 
The well-known tearing instability is an example of such instability 
of a magnetohydrodynamic (MHD) nature \citep{1963PhFl....6..459F}. 
Classical tearing instability leads to fragmentation of the current layer along the direction of the current. 
This fragmentation makes a breach in the shielding created by the current layer 
from which the fast reconnection process can begin \citep{1993SSRv...65..253S}. 
However, it is unable to explain the inhomogeneity in the energy release 
along the direction of the current in the layer observed in solar flares \citep{2015SoPh..290.2909R}. 
There are modifications of the tearing instability that provide oblique fragmentation of the layer, 
designed to describe such inhomogeneity \citep{2012SoPh..277..283A}.

The thermal instability of the plasma is another type of instability \citep{1965ApJ...142..531F}.
It can also explain the inhomogeneity of energy release in solar flares \citep{1982SoPh...75..237S}.
We have found an instability of thermal nature in the piecewise homogeneous
model of the preflare current layer \citep{2021SoPh..296...74L}. 
Periodic fragmentation of the current layer across the direction of the current is the result of the instability. 
We assume that such instability can cause the onset of the fast reconnection phase in the current layer. 
As a result, a quasiperiodic distribution of regions of intense energy release along the current layer is formed. 
The found spatial period of the instability is in the range of 1--10 Mm 
for a wide range of assumed parameters of the coronal plasma. 
This scale is consistent with the observed loop distance in solar flare arcades.

We have also investigated the influence of the guide field on the simulation results \citep{2021SoPh..296...93L}. 
The guide magnetic field is located along the current both outside and inside the current layer. 
The guide field does not participate in reconnection. 
However, it affects the pressure balance at the boundary of the current layer 
and leads to an anisotropy of the transfer coefficients within the layer. 
We have shown that a weak guide field, suppressing the thermal conductivity inside the current layer, 
promotes the formation of the instability, but a strong field leads to its stabilization. 
The spatial scale of the instability remains the same in the expected temperature range of the current layer.

The listed results were obtained under the assumption of an infinite width (along $x$-axis)
of the current layer under the condition $\partial / \partial {x}=0$. 
This made it possible to simplify the derivation of the dispersion relation, 
while maintaining the physical content of the model. 
In addition, it also avoids the appearance of the well-known tearing instability, 
which leads to a similar fragmentation of the current layer along the direction of the current. 
Physically, the condition $\partial / \partial {x}=0$ means that we consider perturbations propagating 
only parallel to the direction of the current in the layer. 
This is the direction we expect to see in the observations,
but it is a prescribed condition and not the result of the model itself. 

In this article, we want to drop the condition $\partial / \partial {x}=0$ and study how the simulation results change. 
In Section~\ref{sec2}, we consider an infinitely wide current layer with perturbations propagating in its plane. 
In Section~\ref{sec3}, we study the dispersion equation properties.
In Section~\ref{sec4}, we calculate the spatial scales of the instability for coronal plasma parameters.
Conclusions are given in Section~\ref{sec5}.

\section{Current Layer Model}
\label{sec2}

\begin{figure}
\begin{center}
\vspace{3mm}
\includegraphics*[width=0.8\linewidth]{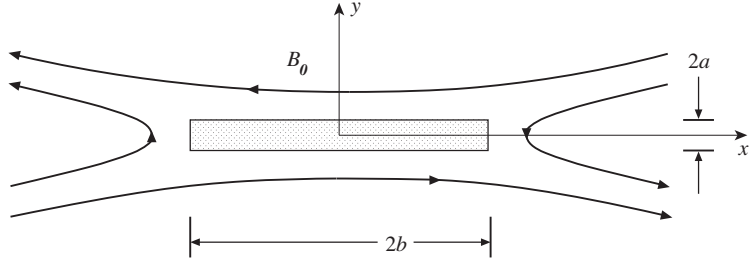}
\end{center}
\caption{Location of the current layer in the coordinate system.
$a$ is the half-thickness, $b$ is the half-width of the layer.}
\label{fig1}
\end{figure}

We consider a piecewise homogeneous MHD model of the current layer. 
The model consists of a current layer located in the $(x,y)$ plane and surrounding plasma (Figure \ref{fig1}). 
The current layer has a half-thickness $a$, half-width $b$, temperature $T_s$, and density $n_s$. 
We consider the current layer to be thin ($b\gg a$), and henceforth take $b \to \infty$.
$T_0$ and $n_0$ are the temperature and concentration of the surrounding plasma. 
The magnetic field $B_0$ in the unperturbed state is present only outside the current layer. 
However, it is possible for a perturbation of the magnetic field to penetrate into the layer. 
The magnetic field is directed against the $x$-axis for positive $y$ and along the $x$-axis for negative $y$. 
Plasma outside the current layer is considered in the ideal MHD approximation. 
The effects of electrical and thermal conductivity as well as viscosity are taken into account inside the layer.
%%% Fig 1 %%%

Plasma behavior is described by the following set of MHD equations \citep{1958ForPh...6..437S, 2006ASSL..340.....S}:
\begin{displaymath}
    \frac{\partial n}{\partial t} + \div \, (n \vec{v})
    = 0 \, ,
\end{displaymath}
\begin{displaymath}
    \mu n \, \frac{\rm{d} \vec{v}}{{\rm d} t}
  = - \grad (2 n k_{_{ \rm B }} T)
  - \frac{1}{4\pi} \, ( \vec{B} \times \cur \vec{B} ) + \eta \, \Delta \vec{v} + \nu \, \grad \, \div \, \vec{v} \, ,
\end{displaymath}
\begin{displaymath}
    \frac{2 n k_{_{ \rm B }} }{\gamma -1} \,
    \frac{{\rm d} T}{{\rm d} t}
  - 2 k_{_{ \rm B }} T \, \frac{ {\rm d} n }{ {\rm d} t } \\
  = \frac{ c^2 }{ ( 4 \pi )^2 \sigma } \, ( {\cur} {\vec B} )^2 + \pder{}{r_\alpha} (\sigma_{\alpha \beta} v_\beta)
  + {\div} \, ( \kappa \grad T ) - \lambda \, (n,T) \, ,
  \end{displaymath}
\begin{displaymath}
    \frac{\partial {\vec B}}{\partial t}
  = {\cur} \, ( {\vec v} \times {\vec B} )
  - \frac{ c^2 }{ 4 \pi } \, {\cur}
    \left( \frac{1}{\sigma} \, {\cur} {\vec B} \right) ,
\end{displaymath}
\begin{equation}
    {\div} {\vec B} = 0 \, .
\label{01}
\end{equation}
Here, $ \mu =1.44 \, m_H $, $ m_H $ is the mass of the hydrogen atom,
$ k_{_{ \rm B }} $ is the Boltzmann constant.
The heat capacity ratio is assumed $\gamma = 5/3$ for simplicity. 
$T$ is the temperature, $n$ is the plasma density, $v$ is the plasma velocity, and $B$ is the magnetic field.
The plasma is assumed to be ideal outside the current layer, 
while dissipative effects are important inside it: 
the finite electrical $\sigma$ and thermal $\kappa$ conductivity of the plasma, 
the viscosity ratios $ \eta $, and $ \nu $,
and the viscous stress tensor $\sigma_{\alpha \beta}$, 
as well as its radiative cooling $\lambda$.
Here $\alpha$ and $\beta$ are tensor indices corresponding to the spatial coordinates $x$, $y$, and $z$.
The radiative cooling function $\lambda \,(n,T)=n^2L(T)$ contains 
the total radiative loss function $L(T)$ that is calculated 
from the CHIANTI 9 atomic database \citep{2019ApJS..241...22D} 
for an optically thin medium with coronal abundance of elements 
(see Figure 1 in \citeauthor{2021SoPh..296...74L}, \citeyear{2021SoPh..296...74L}).
The solution of the system of equations for small perturbations will be found 
separately outside the layer and inside the layer. 
Then the found solutions will be sewn on the boundary, 
which is a tangential MHD discontinuity.

\subsection{Outside the Current Layer}
\label{sec2.1}

The plasma is at rest $v_0=0$, and the dissipative effects are negligible 
$\sigma \to \infty$, $\kappa=0$, $\lambda=0$ outside the current layer.
The symmetry of the model allows us to consider only the positive $y$.
The solution to the problem of small perturbations is assumed 
to be periodic in the plane of the current layer and decay with distance from the layer:
\begin{displaymath}
    f(y,z,t)
  = f_0
  + f_1(y) \, {\rm exp} \, ( - i \omega t + i k_x x + i k_z z ) \, ,
\end{displaymath}
\begin{displaymath}
    f_1(y)=f_1 \, {\rm exp} \, [ - k_{y1} (y-a) ] \, , 
\end{displaymath}
where perturbation amplitudes are
\begin{displaymath}
    f_1 \equiv \{{v_{x1}, v_{y1}, v_{z1}, n_1, T_1, B_{x1}, B_{y1}, B_{z1}}\} \, ,
\end{displaymath}

The linearized system of Equations \ref{01} takes the form:
\begin{equation}
    i \omega \, n_1
  = i k_x \, n_0 v_{x1} - k_{y1} \, n_0 v_{y1} + i k_z \, n_0 v_{z1} \, ,
    \label{02}
\end{equation}
\begin{equation}
    i \omega \, \mu n_0 v_{x1}
  = i k_{x} \, 2 k_{_{ \rm B }}  (n_0 T_1 + T_0 n_1)  \, ,
    \label{03}
\end{equation}
\begin{equation}
    i \omega \, \mu n_0 v_{y1}
  = - k_{y1} \, 2k_{_{ \rm B }}  (n_0 T_1 + T_0 n_1)
    + k_{y1} \, \frac{ B_0  }{ 4 \pi }\,B_{x1} 
    + i k_{x} \, \frac{ B_0  }{ 4 \pi }\,B_{y1}  \, ,
    \label{04}
\end{equation}
\begin{equation}
    i \omega \, \mu n_0 v_{z1}
  = i k_{z} \, 2k_{_{ \rm B }}  (n_0 T_1 + T_0 n_1)
  - i k_{z} \, \frac{ B_0  }{ 4 \pi } \, B_{x1}
  + i k_{x} \, \frac{ B_0  }{ 4 \pi } \, B_{z1} \, ,
    \label{05}
\end{equation}
\begin{equation}
    ( \gamma - 1 ) \, T_0 n_1 = n_0 T_1 \, ,
    \label{06}
\end{equation}
\begin{equation}
    i \omega \, B_{x1}
  = k_{y1} \, B_0 \, v_{y1} - i k_z \, B_0 \, v_{z1} \, ,
    \label{07}
\end{equation}
\begin{equation}
    i \omega \, B_{y1}
  = i k_x \, B_0 \, v_{y1} \, ,
    \label{08}
\end{equation}
\begin{equation}
    i \omega \, B_{z1}
  = i k_{x} \, B_0 \, v_{z1} \, .
    \label{09}
\end{equation}
The determinant of a homogeneous system of the linear Equations \ref{02}--\ref{09}
must be equal to zero for a nontrivial solution to exist.
Therefore, the dispersion relation 
\begin{equation}
    k_{y1}^2 = k_z^2+\frac{V_A^2 k_x^2 + {\it \Gamma}^{\,2}}
    {\left(\frac{1}{V_S^2} + \frac{k_x^2}{{\it \Gamma}^{\,2}}\right)^{-1}\!\!+V_A^2}
    \label{10}
\end{equation}
must be fulfilled outside the current layer.
Here we have introduced the notation for the sound and Alfv\'en speeds
\begin{displaymath}
    V_S = \sqrt{ \frac{ 2 \gamma k_{_{ \rm B }} T_0 }{ \mu }} \, ,
    \qquad
        V_A = \frac{B_0}{ \sqrt{ 4 \pi n_0 \mu } } \, ,
        \qquad
\end{displaymath}
as well as the increment of the instability ${\it \Gamma} = - i \omega$.

%
%%%%%%%%%%%%%%%%%%%%%%%%%%%%%%%%%%%%%%%%%%%%
%
\subsection{Inside the Current Layer}
\label{sec2.2}

The plasma is also at rest $v_s=0$, but the dissipative effects must be considered inside the current layer.
The solution is also sought in the form of a sum of a constant term and a perturbation
\begin{displaymath}
    f(y,z,t) = f_s
             + f_2(y) \, {\rm exp} \, (-i\omega t + i k_x x + ik_zz) \, .
\end{displaymath}
Following \cite{2021SoPh..296...74L}, we consider the dependence of the perturbation on the coordinate $y$ 
in the form of a hyperbolic sine for odd perturbations in $y$ 
\begin{displaymath}
\left\{
\begin{array}{c}
v_{y2}(y)\\
B_{y2}(y)
\end{array}
\right\} =
\left\{
\begin{array}{c}
v_{y2}\\
B_{y2}
\end{array}
\right\}
{\rm sinh} \, (k_{y2} y) \, 
\end{displaymath}
and a hyperbolic cosine for even perturbations in $y$
\begin{displaymath}
\left\{
\begin{array}{c}
v_{x2}(y)\\
v_{z2}(y)\\
n_2(y)\\
T_2(y)\\
B_{x2}(y)\\
B_{z2}(y)
\end{array}
\right\} =
\left\{
\begin{array}{c}
v_{x2}\\
v_{z2}\\
n_2\\
T_2\\
B_{x2}\\
B_{z2}
\end{array}
\right\}
{\rm cosh} \, (k_{y2} y) \, .
\end{displaymath}
In contrast to \cite{2021SoPh..296...74L, 2021SoPh..296...93L}, 
the amplitudes $v_{x2}$ and $B_{x2}$ are chosen to be even to derive a consistent system of linear equations.

Inside the current layer, the linearized system of Equations \ref{01} takes the form:
\begin{equation}
    i \omega \, n_{2}
  = i k_{x} \, n_s v_{x2} + k_{y2} \, n_s v_{y2} + i k_z \, n_s v_{z2} \, ,
    \label{11}
\end{equation}
\begin{equation}
    i \omega \, \mu n_s v_{x2}
  = i k_{x} \, 2 k_{_{ \rm B }} ( n_s T_2 + T_s n_2 ) 
  + (k_z^2+k_{x}^2-k_{y2}^2) \, \eta v_{x2}
  - i \omega i k_{x} \frac{\nu}{n_s}n_2\, ,
    \label{12}
\end{equation}
\begin{equation}
    i \omega \, \mu n_s v_{y2}
  = k_{y2} \, 2 k_{_{ \rm B }} ( n_s T_2 + T_s n_2 ) 
  + (k_z^2+k_{x}^2-k_{y2}^2) \, \eta v_{y2} 
  - i \omega k_{y2} \frac{\nu}{n_s}n_2\, ,
    \label{13}
\end{equation}
\begin{equation}
        i \omega \, \mu n_s v_{z2}
  = i k_{z} \, 2 k_{_{ \rm B }} ( n_s T_2 + T_s n_2 ) 
  + (k_z^2+k_{x}^2-k_{y2}^2) \, \eta v_{z2} 
  - i \omega i k_{z} \frac{\nu}{n_s}n_2\, ,
    \label{14}
\end{equation}
\begin{equation}
    i \omega \, \frac{ 2 k_{_{ \rm B }} n_s }{ \gamma - 1} \, T_2
  - i \omega \, 2 k_{_{ \rm B }} T_s \, n_2 =  (k_z^2+k_{x}^2-k_{y2}^2) \, \kappa \, T_2
   + \pder{ \lambda }{ T } \, T_2 + \pder{ \lambda }{ n } \, n_2 \, ,
    \label{15}
\end{equation}
\begin{equation}
    i \omega \, B_{x2}
  =  (k_z^2+k_{x}^2-k_{y2}^2) \, \nu_{m} \, B_{x2} \, ,
    \label{16}
\end{equation}
\begin{equation}
    i \omega \, B_{y2}
  = (k_z^2+k_{x}^2-k_{y2}^2) \, \nu_{m} \, B_{y2} \, ,
    \label{17}
\end{equation}
\begin{equation}
    i \omega \, B_{z2}
  =  (k_z^2+k_{x}^2-k_{y2}^2) \, \nu_{m} \, B_{z2} \, .
    \label{18}
\end{equation}

The set of Equations \ref{11}--\ref{18} splits into two sets and, 
as a consequence, has two dispersion relations at once.
Equations \ref{16}--\ref{18} under the assumption of a nonzero magnetic perturbation
give the dispersion relation
\begin{equation}
    k_{y2}^2 = k_z^2 + k_x^2 + \frac{{\it \Gamma}}{\nu_m} \, ,
    \label{20}
\end{equation}
where
\begin{displaymath}
    \nu_m
  =  \frac{ c^2 }{4 \pi \sigma} \, 
\end{displaymath}
is the magnetic viscosity.

Using Equation \ref{20} we can eliminate the wave numbers $k_x$, $k_{y2}$, and $k_z$ 
from Equations \ref{11}--\ref{15} and determine the increment of the instability
\begin{equation}
   {\it \Gamma} =
    \frac{ 2 }{ 5 } \, \frac{ \beta-\alpha }{ \tau_\lambda } \, .
    \label{21}
\end{equation}
For details, see the derivation of Equation 33 in \cite{2021SoPh..296...74L}.
The notations for the logarithmic derivatives of the cooling function 
\begin{displaymath}
    \alpha
  = \pder{ {\rm \, ln} \, \lambda }{ {\rm \, ln} \, T } \, ,
  \qquad
  \beta
  = \pder{ {\rm \, ln} \, \lambda }{ {\rm \, ln} \, n } \, 
\end{displaymath}
and characteristic time of the radiative cooling 
\begin{displaymath}
    \tau_\lambda
  =  \frac{ 2 k_{_{ \rm B }} T_s n_s }{ \lambda } \, 
\end{displaymath}
are introduced in Equation \ref{21}.

%
%%%%%%%%%%%%%%%%%%%%%%%%%%%%%%%%%%%%%%%%%%%%
%
\subsection{Boundary of the Current Layer}
\label{sec2.3}

The tangential discontinuity is located at the boundary of the current layer.
Zero plasma velocity ($v_0=0, v_s=0$) and the absence of a component of the magnetic field 
normal to the discontinuity surface indicate this \citep{2015PhyU...58..107L}.
The boundary condition for the tangential discontinuity is that 
the total gas and magnetic pressures on both sides of the discontinuity are equal \citep{1956TrFIAN...13..64S}
\begin{equation}
    2 k_{_{ \rm B }} n_0 T_0 + \frac{ B_0^2}{ 8 \pi } 
  = 2 k_{_{ \rm B }} n_s T_s \, .
    \label{22}
\end{equation}
In addition, velocity perturbations $v_x, v_y, v_z$ will lead 
to a wave-like curvature of the discontinuity surface \citep{2021SoPh..296...74L}.
The linearized boundary conditions are written as follows:
\begin{equation}
    n_0 T_1 + T_0 n_1 - \frac{ B_0 B_{x1} }{ 8 \pi k_{_{ \rm B }} } 
  = ( n_s T_2 + T_s n_2 ) \, {\rm cosh} \, ( k_{y2} a ) \, ,
    \label{23}
\end{equation}
\begin{equation}
    v_{y1} = \pm \, v_{y2} \, {\rm sinh} \, ( k_{y2} a ) \, .
    \label{24}
\end{equation}
We express Equation \ref{23} in terms of perturbation $v_{y1}$ and $v_{y2}$ using 
Equations \ref{04},  \ref{08}, and \ref{11}--\ref{14}.
Then we divide Equation \ref{24} by Equation \ref{23} 
\begin{equation}
    \pm \, \frac{\tau_\nu}{\tau_\sigma} \frac{n_s}{n_0} \, k_{y1}=
    \left(1+ V_A^2 \frac{k_x^2}{{\it \Gamma}^{\,2}}\right) k_{y2} \, {\rm tanh} \, ( k_{y2} a ) \, ,
    \label{25}
\end{equation}
where
\begin{displaymath}
    \frac{\tau_\nu}{\tau_\sigma} =
    1-\frac{ \eta + \nu }{ \mu n_s\, \nu_m } \, .
\end{displaymath}
For details about characteristic times $\tau_\nu$ and $\tau_\sigma$, 
see Equation 9 in \cite{2021SoPh..296...74L}.
Wave numbers $k_{y1}$ and $k_{y2}$ can be eliminated from Equation \ref{25} by using
Equations \ref{10} and \ref{20}, respectively
\begin{eqnarray}
    &\left(\frac{\tau_\nu}{\tau_\sigma} \frac{n_s}{n_0} \right)^2
     \left(k_z^2+\frac{V_A^2 k_x^2 + {\it \Gamma}^{\,2}}
    {\left(\frac{1}{V_S^2} + \frac{k_x^2}{{\it \Gamma}^{\,2}}\right)^{-1}\!\!+V_A^2}\right)\nonumber \\
    &=\left(1+ V_A^2 \frac{k_x^2}{{\it \Gamma}^{\,2}}\right)^2 
    \left(k_z^2 + k_x^2 + \frac{{\it \Gamma}}{\nu_m}\right) \, 
    {\rm tanh}^2 \, \left[ \left(k_z^2 + k_x^2 + \frac{{\it \Gamma}}{\nu_m}\right)^{\!1/2} \!\!a \right] \, .
    \label{26}
\end{eqnarray}
The dispersion Equation \ref{26} is equivalent to the dispersion Equation 32 from \cite{2021SoPh..296...74L}
for $k_x=0$.

Equation \ref{26} has thin and thick approximations depending on the value under the hyperbolic tangent.
\begin{equation}
    k_{z{\rm thin}}^2\simeq\left(\frac{(1+ V_A^2 \frac{k_x^2}{{\it \Gamma}^{\,2}})(k_z^2 + k_x^2 + \frac{{\it \Gamma}}{\nu_m})a}
    {\frac{\tau_\nu}{\tau_\sigma} \frac{n_s}{n_0}}\right)^{\!2}\!\!-\frac{V_A^2 k_x^2 + {\it \Gamma}^{\,2}}
    {\left(\frac{1}{V_S^2} + \frac{k_x^2}{{\it \Gamma}^{\,2}}\right)^{-1}\!\!+V_A^2} \, 
    \label{27}
\end{equation}
for $ {\rm tanh}\,x \simeq x $ and
\begin{equation}
    k_{z{\rm thick}}^2\simeq\frac{(1+ V_A^2 \frac{k_x^2}{{\it \Gamma}^{\,2}})^2(k_z^2 + k_x^2 + \frac{{\it \Gamma}}{\nu_m})}
    {\left(\frac{\tau_\nu}{\tau_\sigma} \frac{n_s}{n_0}\right)^2}-\frac{V_A^2 k_x^2 + {\it \Gamma}^{\,2}}
    {\left(\frac{1}{V_S^2} + \frac{k_x^2}{{\it \Gamma}^{\,2}}\right)^{-1}\!\!+V_A^2} \, 
    \label{28}
\end{equation}
for $ {\rm tanh}\,x \simeq 1 $.
If we use the assumption ${{\it \Gamma}}/{\nu_m} \gg k^2$ which
was fulfilled for coronal plasma with great accuracy in \cite{2021SoPh..296...74L} and
neglect viscosity and the second terms in the Equations \ref{27} and \ref{28}, we find
\begin{equation}
    k_{z{\rm thin}}\simeq\left(1- \frac{V_A^2}{V_x^2}\right)\frac{n_0}{n_s}\,\frac{{\it \Gamma}}{V_D} \, ,
    \label{29}
\end{equation}
\begin{equation}
    k_{z{\rm thick}}\simeq\left(1- \frac{V_A^2}{V_x^2}\right)\frac{n_0}{n_s}\sqrt{\frac{{\it \Gamma}}{\nu_m}} \, ,
    \label{30}
\end{equation}
where $V_D=\nu_m / a$ is the drift velocity and
$V_x^2=-({\it \Gamma}/k_x)^2$ is the square of the $x$-component of the phase velocity of the perturbation.
It will be shown in Section \ref{sec4} that Equations \ref{29} and \ref{30} 
are good simple approximations of the exact dispersion relation for the coronal plasma.

\section{Dispersion Equation Properties}
\label{sec3}

The dispersion Equation \ref{26} connects $k_x$, $k_z$, and ${\it \Gamma}$. 
The instability increment ${\it \Gamma}$ is independently determined by Equation \ref{21}. 
However, $k_z$ can be determined not for every $k_x$ from Equation \ref{26} for a given ${\it \Gamma}$. 
We will demonstrate this as follows. Let us introduce the function
\begin{eqnarray}
    &g({\it \Gamma}, k_x, k_z) = \left(\frac{\tau_\nu}{\tau_\sigma} \frac{n_s}{n_0} \right)^2
     \left(k_z^2+\frac{V_A^2 k_x^2 + {\it \Gamma}^{\,2}}
    {\left(\frac{1}{V_S^2} + \frac{k_x^2}{{\it \Gamma}^{\,2}}\right)^{-1}\!\!+V_A^2}\right)\nonumber\\
    &-\left(1+ V_A^2 \frac{k_x^2}{{\it \Gamma}^{\,2}}\right)^2 
    \left(k_z^2 + k_x^2 + \frac{{\it \Gamma}}{\nu_m}\right) \, 
    {\rm tanh}^2 \, \left[ \left(k_z^2 + k_x^2 + \frac{{\it \Gamma}}{\nu_m}\right)^{\!1/2} \!\!a \right] \, . \nonumber
\end{eqnarray}
The dispersion Equation \ref{26} defines the zeros of this function.
If the function $g$ has no zeros for some given ${\it \Gamma}$ and $k_x$, 
then the dispersion Equation \ref{26} has no corresponding solutions.
We use characteristic conditions of the coronal plasma of an active region 
$n_0=10^{10}$ cm$^{-3}$, $n_s=10^{11}$ cm$^{-3}$, $T_0=10^6$ K, 
$T_s=10^7$ K, $a=10^5$ cm, $\sigma=10^{11}$ s$^{-1}$ \citep{2006ASSL..341.....S}.
The magnetic field $B_0$ is determined by Equation \ref{22}
and is usually about 100 G,
${\it \Gamma}$ is determined by Equation \ref{21}. 

\begin{figure}
\vspace{2mm}
\begin{center}
\includegraphics*[width=\linewidth]{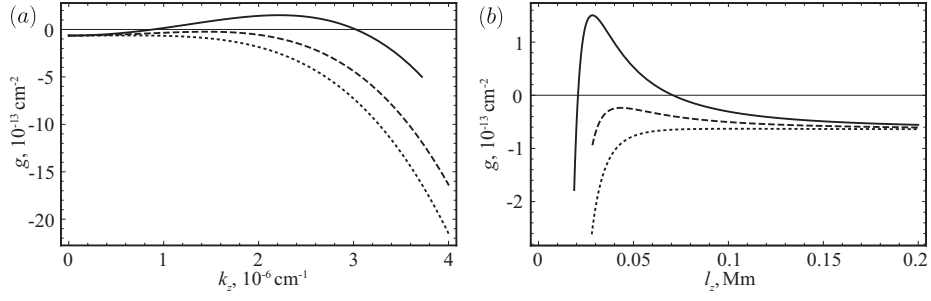}
\end{center}
\caption{$g$ as a function (a) of the wave number $k_z$
and (b) of the spatial period $l_z$ for three values of the wave number $k_x$:
$10^{-9.15}$ cm$^{-2}$ (dotted lines), $10^{-9.20}$ cm$^{-2}$ (dashed lines), $10^{-9.25}$ cm$^{-2}$ (solid lines).
}
\label{fig2}
\end{figure}

Figure \ref{fig2} shows the dependence of the function $g$ on the wave number $k_z$ (Figure \ref{fig2}a)
and on the spatial period $l_z = 2 \pi/ k_z$ (Figure \ref{fig2}b) for three values of the wave number $k_x$:
$10^{-9.15}$ cm$^{-2}$ (dotted lines), $10^{-9.20}$ cm$^{-2}$ (dashed lines), $10^{-9.25}$ cm$^{-2}$ (solid lines).
There is a value $k_{x{\rm max}}$ at which the function $g$ touches the $k_z$-axis . 
The function $g$ has zeros for $k_x \leq k_{x{\rm max}}$ and does not have zeros for $k_x > k_{x{\rm max}}$.
Thus, the dispersion Equation \ref{26} has solutions only for $k_x \leq k_{x{\rm max}}$ for each given ${\it \Gamma}$.
The function $g$ has two zeros for $k_x < k_{x{\rm max}}$. 
With decreasing $k_x$, the lower zero in Figure \ref{fig2}a 
tends to the value corresponding to the dispersion Equation 32 from \cite{2021SoPh..296...74L}, 
while the larger zero quickly tends to infinity and disappears for
\begin{displaymath}
\frac{\tau_\nu}{\tau_\sigma} \frac{n_s}{n_0} >1+ V_A^2 \frac{k_x^2}{{\it \Gamma}^{\,2}} \, .
\end{displaymath}
For this reason, in what follows we will consider solutions of dispersion equations 
corresponding only to the lower zero of the function $g$.
%%% Fig 2 %%%

Let us find the maximum wave numbers $k_{x{\rm max}}$ and  $k_{z{\rm max}}$.
They correspond to the contact of the function $g(k_z)$ and $k_z$-axis.
The maximum point of the function $g(k_z)$ is determined by the equation
\begin{eqnarray}
    \left(\frac{\frac{\tau_\nu}{\tau_\sigma} \frac{n_s}{n_0}}{1+ V_A^2 \frac{k_x^2}{{\it \Gamma}^{\,2}}} \right)^2&=&
   {\rm tanh}^2 \, \left[ \left(k_z^2 + k_x^2 + \frac{{\it \Gamma}}{\nu_m}\right)^{\!1/2} \!\!a \right]\nonumber\\
    &+& a\,(k_z^2 + k_x^2 + \frac{{\it \Gamma}}{\nu_m})^{\!1/2}\,
    \frac{{\rm tanh} \, \left[ \left(k_z^2 + k_x^2 + \frac{{\it \Gamma}}{\nu_m}\right)^{\!1/2} \!\!a \right]}
    {{\rm cosh}^2 \, \left[ \left(k_z^2 + k_x^2 + \frac{{\it \Gamma}}{\nu_m}\right)^{\!1/2} \!\!a \right]} \, .
    \label{31}
\end{eqnarray}
Figure \ref{fig2}a shows that $k_{z{\rm max}} \gg k_{x{\rm max}}$.
Then in the thin approximation, we have from Equations \ref{27} and \ref{31}
\begin{displaymath}
    \left(\frac{\frac{\tau_\nu}{\tau_\sigma} \frac{n_s}{n_0}}{1+ V_A^2 \frac{k_{x{\rm thin}}^2}{{\it \Gamma}^{\,2}}} \right)^2 \simeq
    \frac{a^2}{k_{z{\rm thin}}^2}\left(k_{z{\rm thin}}^2 + \frac{{\it \Gamma}}{\nu_m}\right)^{\!2} \, ,
\end{displaymath}
\begin{displaymath}
    \left(\frac{\frac{\tau_\nu}{\tau_\sigma} \frac{n_s}{n_0}}{1+ V_A^2 \frac{k_{x{\rm thin}}^2}{{\it \Gamma}^{\,2}}} \right)^2 \simeq
    2 a^2\left(k_{z{\rm thin}}^2 + \frac{{\it \Gamma}}{\nu_m}\right) \, ,
\end{displaymath}
respectively. From here
\begin{equation}
    k_{z{\rm thin max}}^2 \simeq \frac{{\it \Gamma}}{\nu_m} \, , \qquad     k_{x{\rm thin max}}^2 \simeq \left(\frac{\tau_\nu}{\tau_\sigma} \, \frac{n_s}{n_0} \, \frac{1}{2 a} \sqrt{\frac{\nu_m}{{\it \Gamma}}}-1\right) 
    \left(\frac{{\it \Gamma}}{V_A}\right)^{\!2} \, .
    \label{32}
\end{equation}
In the thick approximation, we have from Equations \ref{28} and \ref{31}
\begin{displaymath}
    \left(\frac{\frac{\tau_\nu}{\tau_\sigma} \frac{n_s}{n_0}}{1+ V_A^2 \frac{k_{x{\rm thick}}^2}{{\it \Gamma}^{\,2}}} \right)^2 \simeq
    \frac{1}{k_{z{\rm thick}}^2}\left(k_{z{\rm thick}}^2 + \frac{{\it \Gamma}}{\nu_m}\right) \, ,
\end{displaymath}
\begin{displaymath}
    \left(\frac{\frac{\tau_\nu}{\tau_\sigma} \frac{n_s}{n_0}}{1+ V_A^2 \frac{k_{x{\rm thick}}^2}{{\it \Gamma}^{\,2}}} \right)^2 \simeq
    1 \, ,
\end{displaymath}
respectively, and
\begin{equation}
    k_{z{\rm thick max}}^2 \gg \frac{{\it \Gamma}}{\nu_m} \, , \qquad     k_{x{\rm thick max}}^2 \simeq \left(\frac{\tau_\nu}{\tau_\sigma} \, \frac{n_s}{n_0}-1\right) 
    \left(\frac{{\it \Gamma}}{V_A}\right)^{\!2} \, .
    \label{33}
\end{equation}
We use Equations \ref{32} and \ref{33} in Section \ref{sec4} to compare the spatial scales of instability
$l_x = 2 \pi/ k_x$ and $l_z = 2 \pi/ k_z$.

\section{Spatial Scales of the Instability}
\label{sec4}

\begin{figure}
\vspace{2mm}
\begin{center}
\includegraphics*[width=\linewidth]{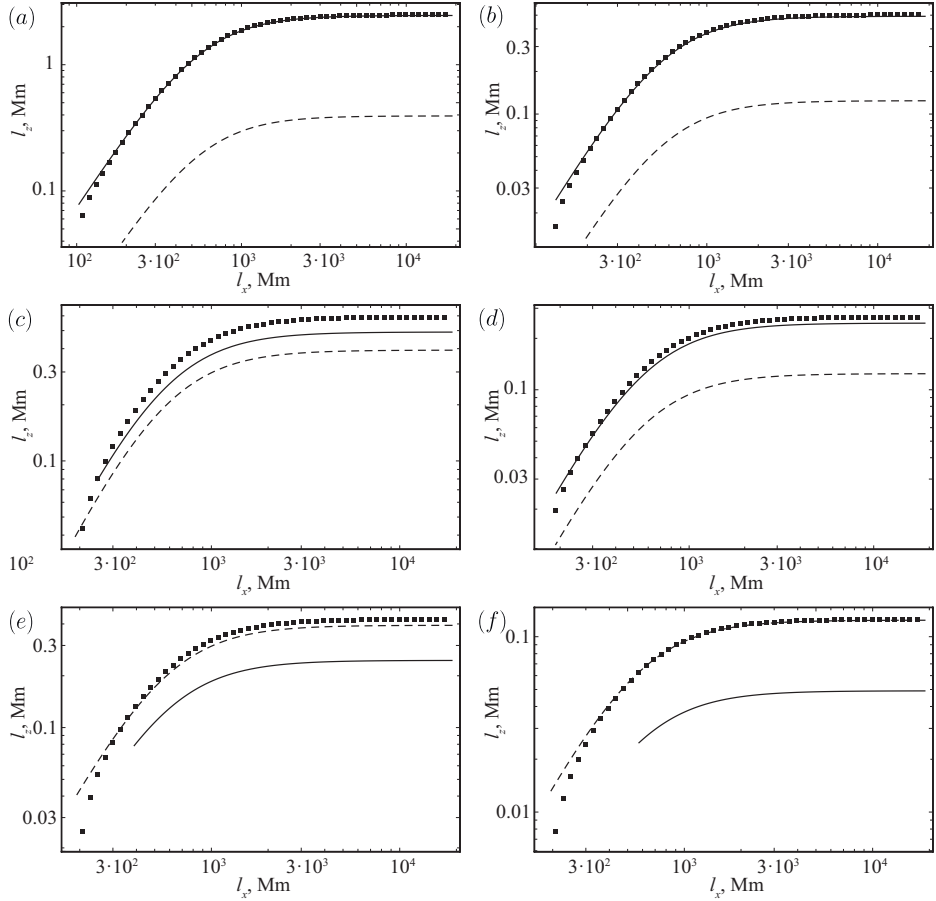}
\end{center}
\caption{The relationship between the spatial scales of the instability $l_x$ and $l_z$
calculated from Equation \ref{26} with $n_0=10^{10}$ and $n_s=10^{11}$ (squares).
Thin (Equations \ref{29}) and thick (Equations \ref{30}) approximations 
are shown with solid and dashed lines, respectively.
Parameters of the current layer: 
(a)~$a=10^{5} {\rm \, cm}$,  
$\sigma=10^{11} {\rm \, s}^{-1}$,
(b)~$a=5\times10^{4} {\rm \, cm}$,  
$\sigma=10^{12} {\rm \, s}^{-1}$,
(c)~$a=5\times10^{5} {\rm \, cm}$,  
$\sigma=10^{11} {\rm \, s}^{-1}$,
(d)~$a=10^{5} {\rm \, cm}$,  
$\sigma=10^{12} {\rm \, s}^{-1}$,
(e)~$a=10^{6} {\rm \, cm}$,  
$\sigma=10^{11} {\rm \, s}^{-1}$,
(f)~$a=5\times10^{5} {\rm \, cm}$,  
$\sigma=10^{12} {\rm \, s}^{-1}$.
}
\label{fig3}
\end{figure}

\begin{figure}
\vspace{2mm}
\begin{center}
\includegraphics*[width=\linewidth]{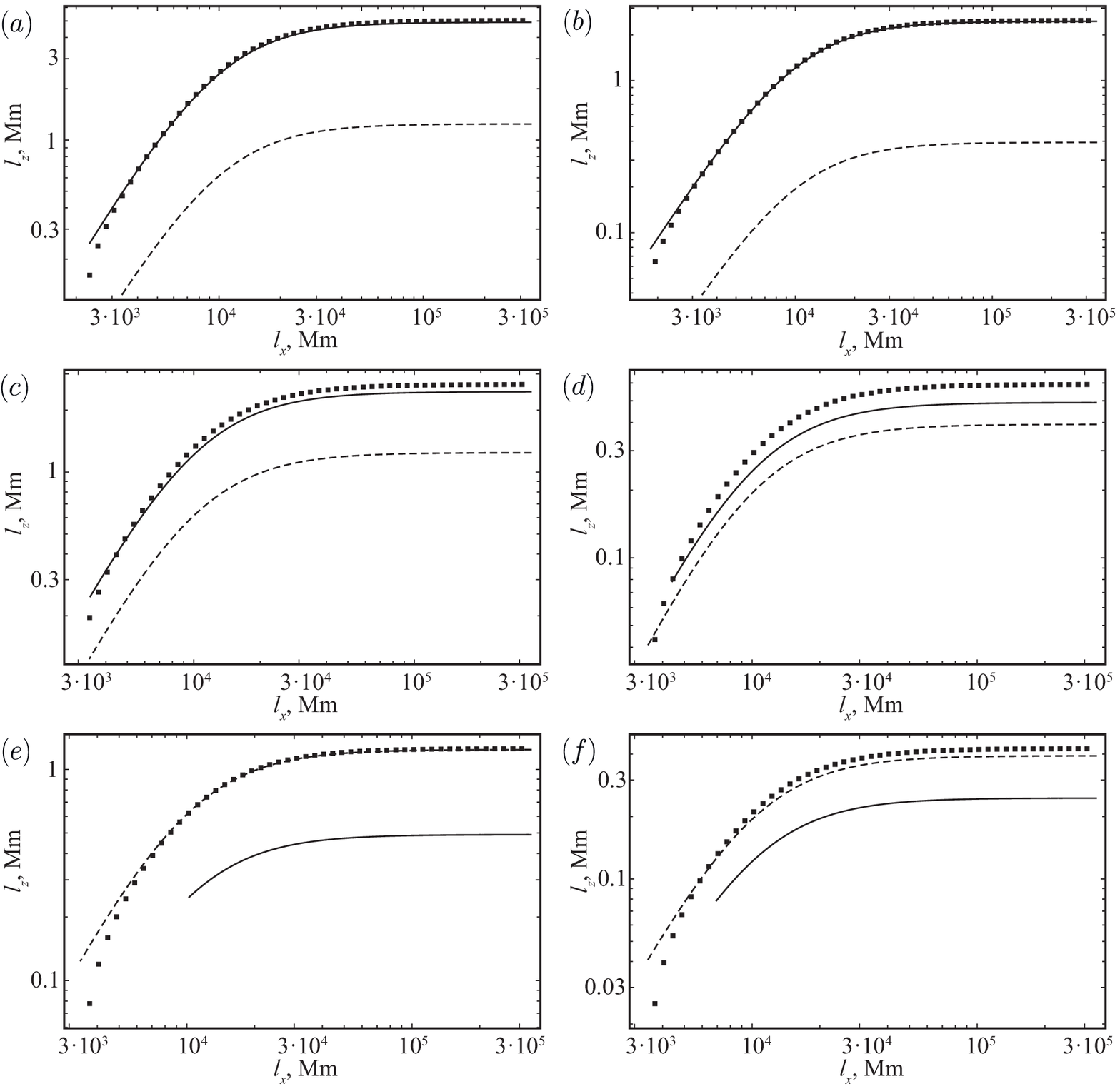}
\end{center}
\caption{The relationship between the spatial scales of the instability $l_x$ and $l_z$
calculated from Equation \ref{26} with $n_0=10^{9}$ and $n_s=10^{10}$ (squares).
Thin (Equations \ref{29}) and thick (Equations \ref{30}) approximations 
are shown with solid and dashed lines, respectively.
Parameters of the current layer: 
(a)~$a=5\times10^{5} {\rm \, cm}$,  
$\sigma=10^{11} {\rm \, s}^{-1}$,
(b)~$a=10^{5} {\rm \, cm}$,  
$\sigma=10^{12} {\rm \, s}^{-1}$,
(c)~$a=10^{6} {\rm \, cm}$,  
$\sigma=10^{11} {\rm \, s}^{-1}$,
(d)~$a=5\times10^{5} {\rm \, cm}$,  
$\sigma=10^{12} {\rm \, s}^{-1}$,
(e)~$a=5\times10^{6} {\rm \, cm}$,  
$\sigma=10^{11} {\rm \, s}^{-1}$,
(f)~$a=10^{6} {\rm \, cm}$,  
$\sigma=10^{12} {\rm \, s}^{-1}$.
}
\label{fig4}
\end{figure}

The spatial scales of the instability $l_x = 2 \pi/ k_x$ and $l_z = 2 \pi/ k_z$ 
are determined by the dispersion Equation \ref{26} and
its simple approximations (Equations \ref{29} and \ref{30}).
Figures \ref{fig3} and \ref{fig4} show the relationship between the spatial scales of the instability 
calculated from Equation \ref{26} (squares).
Thin (Equation \ref{29}) and thick (Equation \ref{30}) approximations 
are also shown in Figures \ref{fig3} and \ref{fig4} with solid and dashed lines, respectively.
The temperature of the surrounding plasma $T_0=10^6$ K and 
the temperature of the current layer $T_s=10^7$ K were used for calculations.
Figures \ref{fig3} and \ref{fig4} show calculation results for concentrations of the surrounding plasma 
$n_0=10^{10}$ cm$^{-3}$ and $n_0=10^{9}$ cm$^{-3}$, respectively.
The density contrast is the same $n_s/n_0=10$ for all calculations.
The parameters of the current layer $a$ and $\sigma$ are indicated in the figure caption.
The increment of the instability ${\it \Gamma}$ and the magnetic field strength $B_0$ were
calculated from Equations \ref{21} and \ref{22}, respectively.
The minimum values for the thin and thick approximations are taken from Equations \ref{32} and \ref{33}, respectively.
%%% Fig 3 %%%
%%% Fig 4 %%%

Figures \ref{fig3} and \ref{fig4} have several features.

1) All squares in Figures \ref{fig3} and \ref{fig4} go out to the value of the spatial scale $l_z$ 
calculated in the model with $\partial / \partial {x}=0$ \citep{2021SoPh..296...74L}
when the spatial scale $l_x$ increases. 
The spatial scale $l_z$ decreases at small $l_x$.
However, the reduction in the scale $l_z$ is approximately an order of magnitude for the minimal scale $l_x$.
This indicates a weak influence of the oblique fragmentation of the current layer on the model results.

2) Squares in Figures \ref{fig3} and \ref{fig4} are in good agreement with the thin or thick approximation, 
depending on the selected half-thickness $a$ of the current layer. 
The largest discrepancies occur at the minimum values $l_z$. 
This is due to the fact that in these values the assumption ${{\it \Gamma}}/{\nu_m} \gg k^2$
used to derive Equations \ref{29} and \ref{30} is not correct
(see Equations \ref{32} and \ref{33}).

3) Figures \ref{fig3} and \ref{fig4} allow us to determine the maximum angle of propagation of the perturbation 
in relation to the direction of the current in the layer. 
It is equal to the angle of inclination of the tangent to the graph of the function $l_z(l_x)$ passing through the origin.
For the dependencies shown in Figures \ref{fig3} and \ref{fig4}, it does not exceed $0.2^\circ$ 
and decreases with increasing layer thickness and plasma conductivity
and decreasing concentration of the surrounding plasma.
This value can increase for other realistic parameters of the coronal plasma 
due to an increase in the instability scale for extremely thin current layers ($a\approx10^4$~cm), 
but not more than 10 times (up to $2^\circ$). 
Hence, it follows that fragmentation transverse to the current 
is a natural property of the thermal instability of the current layer model under consideration.

\section{Conclusion}
\label{sec5}

Earlier, we studied the stability of piecewise homogeneous current layer models
\citep{2021SoPh..296...74L, 2021SoPh..296...93L}
with respect to small perturbations and discovered an instability of thermal nature. 
The instability resulted in fragmentation of the current layer across the direction of the current. 
However, the direction of fragmentation was determined by the 
search for a wave solution propagating along the current. 
In this article, we remove the restriction on the direction of the wave solution. 
We estimate the possibility of an oblique fragmentation formation as a result of the thermal instability of the current layer.

We found the dispersion Equation \ref{26} for the thermal instability of the infinitely wide current layer 
and its simple approximations (Equations \ref{29} and \ref{30}). 
It is shown that the dispersion Equation \ref{26} has no solutions 
for sufficiently small spatial periods of the instability. 
Equations \ref{32} and \ref{33} estimate the maximum wave numbers of the instability.

The influence of the oblique propagation of the perturbation on the model results 
is generally insignificant, but extreme changes can reach an order of magnitude (Figures \ref{fig3} and \ref{fig4}). 
Thus, oblique fragmentation can lead to a decrease in the estimate of the spatial period 
of localization of elementary energy release in solar flares to 0.1--1 Mm instead of 1--10 Mm obtained earlier.

We have established that fragmentation transverse to the current is a natural feature of the model. 
The maximum deviation of the wave vector of the perturbation from the direction of the current 
does not exceed $0.2^\circ$ for the considered examples. 
This value can increase for extremely thin current layers, but not more than $2^\circ$
for the realistic parameters of the coronal plasma.

\vspace{1cm}
\noindent{\bf Disclosure of Potential Conflicts of Interest} The author declares that there are no conflicts of interest.

     % format of references provided by the journal (.bst)
\bibliographystyle{spr-mp-sola}
     % name your Bibtex file containing your references (.bib)
\bibliography{sola_bibliography}  

\end{article} 

\end{document}